\documentclass{aa}

\usepackage{graphicx}
\usepackage{txfonts}
\usepackage{color}
\usepackage{stfloats}
\usepackage{multirow}

\newcommand{\tablenotea}[1]{\parbox{18.2cm}{\indent \footnotesize{#1}}}
\newcommand{\acr}{Acc. Chem. Res.}
\newcommand{\acsesc}{ACS Earth Space Chem.}
\newcommand{\cjc}{Can. J. Chem.}

\newcommand{\fdis}{Faraday Discuss.}

\newcommand{\ijmsip}{Int. J. Mass Spectrom. Ion Process.}

\newcommand{\jms}{J. Mol. Spectr.}
\newcommand{\jmst}{J. Mol. Struct.}

\newcommand{\jpca}{J. Phys. Chem. A}
\newcommand{\jpcrd}{J. Phys. Chem. Ref. Data}
\newcommand{\jppa}{J. Photochem. Photobiol. A}

\newcommand{\molphys}{Mol. Astrophys.}

\newcommand{\physs}{Phys. Scr.}

\newcommand{\rsca}{RSC Adv.}

\newcommand{\sic}{Symposium (International) on Combustion}
\newcommand{\tca}{Theor. Chem. Acc.}
\newcommand{\znat}{Z. Naturforschg.}

\begin{document}

\title{Detection of ethanol, acetone, and propanal in TMC-1:\\ New O-bearing complex organics in cold sources\thanks{Based on observations carried out with the Yebes 40m telescope (projects 19A003, 20A014, 20D023, 21A011, and 21D005). The 40m radio telescope at Yebes Observatory is operated by the Spanish Geographic Institute (IGN; Ministerio de Transportes, Movilidad y Agenda Urbana).}}

\titlerunning{New O-bearing complex organic molecules in \mbox{TMC-1}}
\authorrunning{Ag\'undez et al.}

\author{M.~Ag\'undez\inst{1}, J.-C.~Loison\inst{2}, K. M.~Hickson\inst{2}, V.~Wakelam\inst{3}, R.~Fuentetaja\inst{1}, C.~Cabezas\inst{1}, N.~Marcelino\inst{4,5}, B.~Tercero\inst{4,5}, P.~de~Vicente\inst{5}, \and J.~Cernicharo\inst{1}}

\institute{
Instituto de F\'isica Fundamental, CSIC, Calle Serrano 123, E-28006 Madrid, Spain\\
\email{marcelino.agundez@csic.es, jose.cernicharo@csic.es} \and
Institut des Sciences Mol\'eculaires (ISM), CNRS, Univ. Bordeaux, 351 cours de la Lib\'eration, 33400, Talence, France\\
\email{jean-christophe.loison@u-bordeaux.fr} \and
Laboratoire d'Astrophysique de Bordeaux, Univ. Bordeaux, CNRS, B18N, all\'ee Geoffroy Saint-Hilaire, 33615 Pessac, France \and
Observatorio Astron\'omico Nacional, IGN, Calle Alfonso XII 3, E-28014 Madrid, Spain \and
Observatorio de Yebes, IGN, Cerro de la Palera s/n, E-19141 Yebes, Guadalajara, Spain
}

\date{Received; accepted}

% \abstract{}{}{}{}{} 
% 5 {} token are mandatory
 
\abstract
% context heading (optional), leave it empty if necessary
{We present the detection of ethanol (C$_2$H$_5$OH), acetone (CH$_3$COCH$_3$), and propanal (C$_2$H$_5$CHO) toward the cyanopolyyne peak of \mbox{TMC-1}. These three O-bearing complex organic molecules are known to be present in warm interstellar clouds, but had never been observed in a starless core. The addition of these three new pieces to the puzzle of complex organic molecules in cold interstellar clouds stresses the rich chemical diversity of cold dense cores in stages prior to the onset of star formation. The detections of ethanol, acetone, and propanal were made in the framework of QUIJOTE, a deep line survey of \mbox{TMC-1} in the Q band that is being carried out with the Yebes\,40m telescope. We derive column densities of (1.1\,$\pm$\,0.3)\,$\times$\,10$^{12}$ cm$^{-2}$ for C$_2$H$_5$OH, (1.4\,$\pm$\,0.6)\,$\times$\,10$^{11}$ cm$^{-2}$ for CH$_3$COCH$_3$, and (1.9\,$\pm$\,0.7)\,$\times$\,10$^{11}$ cm$^{-2}$ for C$_2$H$_5$CHO. The formation of these three O-bearing complex organic molecules is investigated with the aid of a detailed chemical model which includes gas and ice chemistry. The calculated abundances at a time around 2\,$\times$\,10$^5$ yr are in reasonable agreement with the values derived from the observations. The formation mechanisms of these molecules in our chemical model are as follows. Ethanol is formed on grains by addition of atomic carbon on methanol followed by hydrogenation and non-thermal desorption. Acetone and propanal are produced by the gas-phase reaction between atomic oxygen and two different isomers of the C$_3$H$_7$ radical, where the latter follows from the hydrogenation of C$_3$ on grains followed by non-thermal desorption. A gas-phase route involving the formation of (CH$_3$)$_2$COH$^+$ through several ion-neutral reactions followed by its dissociative recombination with electrons do also contribute to the formation of acetone.}


\maketitle

\section{Introduction}

Interstellar clouds are known to host a wide variety of complex organic molecules (COMs). Among them there are several O-bearing species with a high degree of hydrogenation, such as dimethyl ether, methyl formate, ethanol, acetone, and acetic acid, that are well known on Earth because they are used in organic chemistry laboratories. These species have been traditionally observed in warm interstellar regions, such as hot cores, hot corinos, and Galactic center clouds \citep{Blake1987,Cazaux2003,Requena-Torres2006}, where they are thought to be formed on dust grains and released to the gas phase upon thermal desorption \citep{Garrod2008}.

In recent years, O-bearing complex organic molecules typical of warm clouds, such as methyl formate (HCOOCH$_3$) and dimethyl ether (CH$_3$OCH$_3$), have been also observed in cold clouds \citep{Oberg2010,Cernicharo2012,Bacmann2012,Taquet2017,Jimenez-Serra2016,Soma2018,Agundez2019,Agundez2021}. These detections came as a surprise because the mechanism responsible for the formation of these molecules in warm sources, which relies on the mobility of heavy radicals on grain surfaces and the thermal desorption of ices, is closed at the very low temperatures of these sources, typically around 10 K. Several scenarios have been proposed (e.g., \citealt{Vasyunin2013,Ruaud2015,Balucani2015,Shingledecker2018,Jin2020}), although there is not yet consensus on which if any is the correct one. Thus, the formation of methyl formate and dimethyl ether in cold sources continues to be an open problem in astrochemistry \citep{Herbst2022}.

In addition to methyl formate and dimethyl ether, two new O-bearing COMs, propenal (C$_2$H$_3$CHO) and vinyl alcohol (C$_2$H$_3$OH), were reported recently in the cold starless core \mbox{TMC-1} \citep{Agundez2021}. Here we report the detection of three new O-bearing COMs toward \mbox{TMC-1}: ethanol (C$_2$H$_5$OH), acetone (CH$_3$COCH$_3$), and propanal (C$_2$H$_5$CHO). These three molecules are well known in warm interstellar clouds \citep{Snyder2002,Hollis2004,Requena-Torres2006,Lykke2017,Manigand2020} but have never been observed in a cold source. We also present a detailed chemical model to investigate the formation of these O-bearing COMs in \mbox{TMC-1}.

\section{Astronomical observations}

\begin{table*}
\small
\caption{Observed line parameters in \mbox{TMC-1}.}
\label{table:lines}
\centering
\begin{tabular}{l@{\hspace{0.9cm}}l@{\hspace{0.55cm}}c@{\hspace{0.55cm}}c@{\hspace{0.55cm}}c@{\hspace{0.55cm}}c@{\hspace{0.55cm}}c@{\hspace{0.55cm}}c@{\hspace{0.55cm}}c@{\hspace{0.55cm}}c}
\hline \hline
Molecule & Transition & $\nu_{\rm calc}$ & $E_{up}$ & $T_{b, \rm \, calc}$\,$^a$ & $V_{\rm LSR}$\,$^b$ & $\Delta v$\,$^b$ & $T_A^*$ peak\,$^b$ & $\int T_A^* dv$\,$^b$  & S/N\,$^c$ \\
         &            & (MHz)      & (K)            & (mK) & (km s$^{-1}$) & (km s$^{-1}$)    & (mK)         & (mK km s$^{-1}$) & ($\sigma$) \\
\hline
C$_2$H$_5$OH   & 4$_{1,3}$-4$_{0,4}$ & 32742.830 &  9.9 & 3.68 & 5.57 $\pm$ 0.12 & 1.03 $\pm$ 0.24 & 0.67 $\pm$ 0.17 & 0.73 $\pm$ 0.16 & 7.2 \\
               & 5$_{1,4}$-5$_{0,5}$ & 36417.242 & 14.3 & 2.31 & 5.83 $\pm$ 0.09 & 1.49 $\pm$ 0.22 & 0.76 $\pm$ 0.12 & 1.21 $\pm$ 0.15 & 14.7 \\
               & 1$_{1,1}$-0$_{0,0}$ & 43026.811 &  2.1 & 4.62 & 5.76 $\pm$ 0.04 & 1.09 $\pm$ 0.07 & 1.90 $\pm$ 0.18 & 2.21 $\pm$ 0.15 & 22.8 \\
               & 4$_{0,4}$-3$_{1,3}$ & 46832.826 &  8.4 & 3.14 & 5.68 $\pm$ 0.03 & 0.90 $\pm$ 0.06 & 2.38 $\pm$ 0.23 & 2.28 $\pm$ 0.15 & 21.1 \\
\hline
CH$_3$COCH$_3$ & 3$_{0,3}$-2$_{1,2}$ $EE$ & 33562.123 & 3.6 & 1.11 & 5.87 $\pm$ 0.14 & 0.99 $\pm$ 0.25 & 0.40 $\pm$ 0.13 & 0.42 $\pm$ 0.11 & 5.6 \\
               & 3$_{0,3}$-2$_{1,2}$ $AA$ & 33566.289 & 3.6 & 0.75 & 5.46 $\pm$ 0.10 & 0.90 $\pm$ 0.16 & 0.46 $\pm$ 0.12 & 0.44 $\pm$ 0.09 & 6.6 \\
               & 3$_{1,3}$-2$_{0,2}$ $EE$ & 34092.973 & 3.6 & 1.15 & 5.89 $\pm$ 0.17 & 1.36 $\pm$ 0.40 & 0.44 $\pm$ 0.15 & 0.64 $\pm$ 0.16 & 6.3 \\
               & 2$_{2,1}$-1$_{1,0}$ $EE$ & 35381.289 & 2.7 & 0.85 & --              & --              & --              & --              & --\,$^d$ \\
               & 4$_{0,4}$-3$_{1,3}$ $EE$ & 43597.127 & 5.7 & 1.54 & 5.95 $\pm$ 0.07 & 0.90 $\pm$ 0.16 & 0.79 $\pm$ 0.17 & 0.76 $\pm$ 0.12 & 9.2 \\
               & 4$_{0,4}$-3$_{1,3}$ $AA$ & 43604.463 & 5.7 & 0.61 & --              & --              & --              & --              & --\,$^e$ \\
               & 4$_{1,4}$-3$_{0,3}$ $AE$ & 43680.468 & 5.8 & 0.61 & --              & --              & --              & --              & --\,$^f$ \\
               & 4$_{1,4}$-3$_{0,3}$ $EE$ & 43689.327 & 5.7 & 1.54 & 5.73 $\pm$ 0.08 & 0.50 $\pm$ 0.17 & 0.61 $\pm$ 0.17 & 0.33 $\pm$ 0.10 & 5.4 \\
               & 4$_{1,4}$-3$_{0,3}$ $AA$ & 43698.504 & 5.7 & 1.03 & 5.84 $\pm$ 0.11 & 0.74 $\pm$ 0.24 & 0.71 $\pm$ 0.21 & 0.55 $\pm$ 0.16 & 6.0 \\
               & 3$_{2,2}$-2$_{1,1}$ $EE$ & 45233.560 & 4.6 & 0.94 & --              & --              & --              & --              & --\,$^e$ \\
\hline
C$_2$H$_5$CHO  & 4$_{0,4}$-3$_{1,3}$ & 32897.820 & 5.0 & 0.72 & --              & --              & --              & --              & --\,$^d$ \\
               & 3$_{1,2}$-2$_{1,1}$ & 33346.830 & 3.8 & 0.96 & 5.59 $\pm$ 0.15 & 0.99 $\pm$ 0.26 & 0.38 $\pm$ 0.10 & 0.40 $\pm$ 0.11 & 6.9 \\
               & 3$_{1,3}$-2$_{0,2}$ & 39061.035 & 3.4 & 1.10 & 5.78 $\pm$ 0.10 & 0.79 $\pm$ 0.20 & 0.45 $\pm$ 0.15 & 0.38 $\pm$ 0.09 & 5.3 \\
               & 4$_{1,4}$-3$_{1,3}$ & 39177.138 & 5.3 & 1.26 & 5.93 $\pm$ 0.10 & 0.73 $\pm$ 0.18 & 0.46 $\pm$ 0.15 & 0.36 $\pm$ 0.09 & 5.2 \\
               & 4$_{0,4}$-3$_{0,3}$ & 40915.768 & 5.0 & 1.47 & 5.81 $\pm$ 0.12 & 1.11 $\pm$ 0.31 & 0.59 $\pm$ 0.16 & 0.70 $\pm$ 0.16 & 7.9 \\
               & 4$_{2,3}$-3$_{2,2}$ & 41883.719 & 7.2 & 0.78 & --              & --              & --              & --              & --\,$^f$ \\
               & 4$_{2,2}$-3$_{2,1}$ & 42936.277 & 7.3 & 0.80 & --              & --              & --              & --              & --\,$^e$ \\
               & 5$_{0,5}$-4$_{1,4}$ & 44189.959 & 7.4 & 0.97 & --              & --              & --              & --              & --\,$^f$ \\
               & 4$_{1,3}$-3$_{1,2}$ & 44319.499 & 5.9 & 1.30 & 6.07 $\pm$ 0.07 & 0.56 $\pm$ 0.19 & 0.67 $\pm$ 0.18 & 0.40 $\pm$ 0.11 & 5.8 \\
               & 4$_{1,4}$-3$_{0,3}$ & 47195.073 & 5.3 & 1.31 & 5.90 $\pm$ 0.11 & 0.89 $\pm$ 0.29 & 0.92 $\pm$ 0.25 & 0.88 $\pm$ 0.22 & 7.6 \\
               & 5$_{1,5}$-4$_{1,4}$ & 48801.725 & 7.6 & 1.40 & 5.63 $\pm$ 0.11 & 0.60 $\pm$ 0.19 & 0.79 $\pm$ 0.28 & 0.51 $\pm$ 0.16 & 4.9 \\
\hline
\end{tabular}
\tablenotea{\\
$^a$\,$T_{b, \rm \, calc}$ is the peak brightness temperature calculated in local thermodynamic equilibrium (LTE) for a rotational temperature of 6.0 K. We adopted the column density and line width derived for each molecule (see Sec.~\ref{sec:results}).\\
$^b$\,The line parameters $V_{\rm LSR}$, $\Delta v$, $T_A^*$ peak, and $\int T_A^* dv$ and the associated errors are derived from a Gaussian fit to each line profile. $\Delta v$ is the full width at half maximum (FWHM).\\
$^c$\,The signal-to-noise ratio is computed as S/N = $\int T_A^* dv$ / [rms $\times$ $\sqrt{\Delta v \times \delta \nu (c / \nu_{\rm calc})}$], where $c$ is the speed of light, $\delta \nu$ is the spectral resolution (0.03815 MHz), the rms is given in the uncertainty of $T_A^*$ peak, and the rest of parameters are given in the table.\\
$^d$\,Line is only marginally detected.\\
$^e$\,Line is not detected. Expected intensity is within the noise level.\\
$^f$\,Line is not detected. It overlaps with a frequency-switching negative artifact.
}
\end{table*}

The data presented here are part of QUIJOTE (Q-band Ultrasensitive Inspection Journey to the Obscure \mbox{TMC-1} Environment; \citealt{Cernicharo2021}), which is an ongoing Q-band line survey carried out with the Yebes\,40m telescope at the position of the cyanopolyyne peak of \mbox{TMC-1} ($\alpha_{J2000}=4^{\rm h} 41^{\rm  m} 41.9^{\rm s}$ and $\delta_{J2000}=+25^\circ 41' 27.0''$). Details on the line survey are given in previous papers of the QUIJOTE series (e.g., \citealt{Cernicharo2021}), while the data reduction procedure adopted is explained in \cite{Cernicharo2022}. Briefly, QUIJOTE uses a 7 mm receiver covering the Q band (31.0-50.3 GHz) connected to a fast Fourier transform spectrometer providing a spectral resolution of 38.15 kHz (see \citealt{Tercero2021}). The data presented here correspond to observations carried out from November 2019 to November 2022, which amount to a total on-source telescope time of 758 h. The frequency-switching technique was used with frequency throws of 8 and 10 MHz. The intensity scale adopted is antenna temperature, $T_A^*$. The estimated uncertainty due to calibration in $T_A^*$ is 10~\%. The antenna temperature can be converted to main beam brightness temperature, $T_{mb}$, by dividing by $B_{\rm eff}$/$F_{\rm eff}$, where $B_{\rm eff}$ and $F_{\rm eff}$ are the beam and forward efficiencies, respectively. For the Yebes 40m telescope in the Q band, $B_{\rm eff}$ can be fitted as a function of frequency as $B_{\rm eff}$\,=\,0.797\,$\exp$[$-$($\nu$(GHz)/71.1)$^2$] using the values measured in 2022 that are reported in the webpage of the Yebes\,40m telescope\footnote{\texttt{https://rt40m.oan.es/rt40m\_en.php}}, while measured values of $F_{\rm eff}$ range from 0.9 to 0.97 and here we adopt $F_{\rm eff}$\,=\,0.97. The half power beam width (HPBW) is given by HPBW($''$)\,=\,1763/$\nu$(GHz). All data were analyzed using GILDAS\footnote{\texttt{http://www.iram.fr/IRAMFR/GILDAS/}}.

\section{Results} \label{sec:results}

\begin{figure}
\centering
\includegraphics[angle=0,width=\columnwidth]{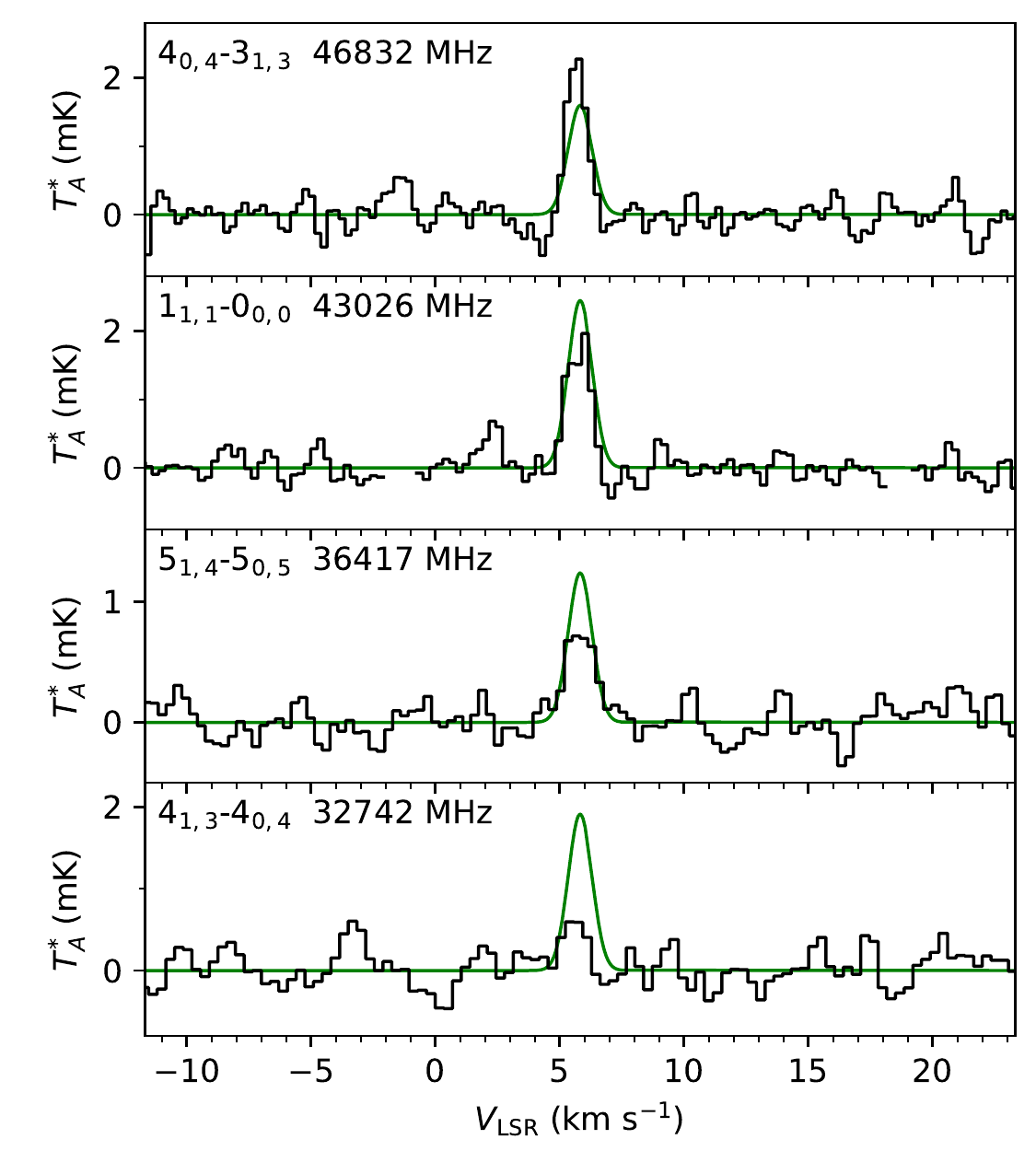}
\caption{Lines of C$_2$H$_5$OH observed in \mbox{TMC-1} (see parameters in Table~\ref{table:lines}). In green we show the calculated spectra for $N$\,=\,1.1\,$\times$\,10$^{12}$\,cm$^{-2}$, $T_{\rm rot}$\,=\,6.0\,K, FWHM\,=\,1.13 km s$^{-1}$, and $\theta_s$\,=\,80\,$''$.} \label{fig:lines_c2h5oh}
\end{figure}

\begin{figure}
\centering
\includegraphics[angle=0,width=\columnwidth]{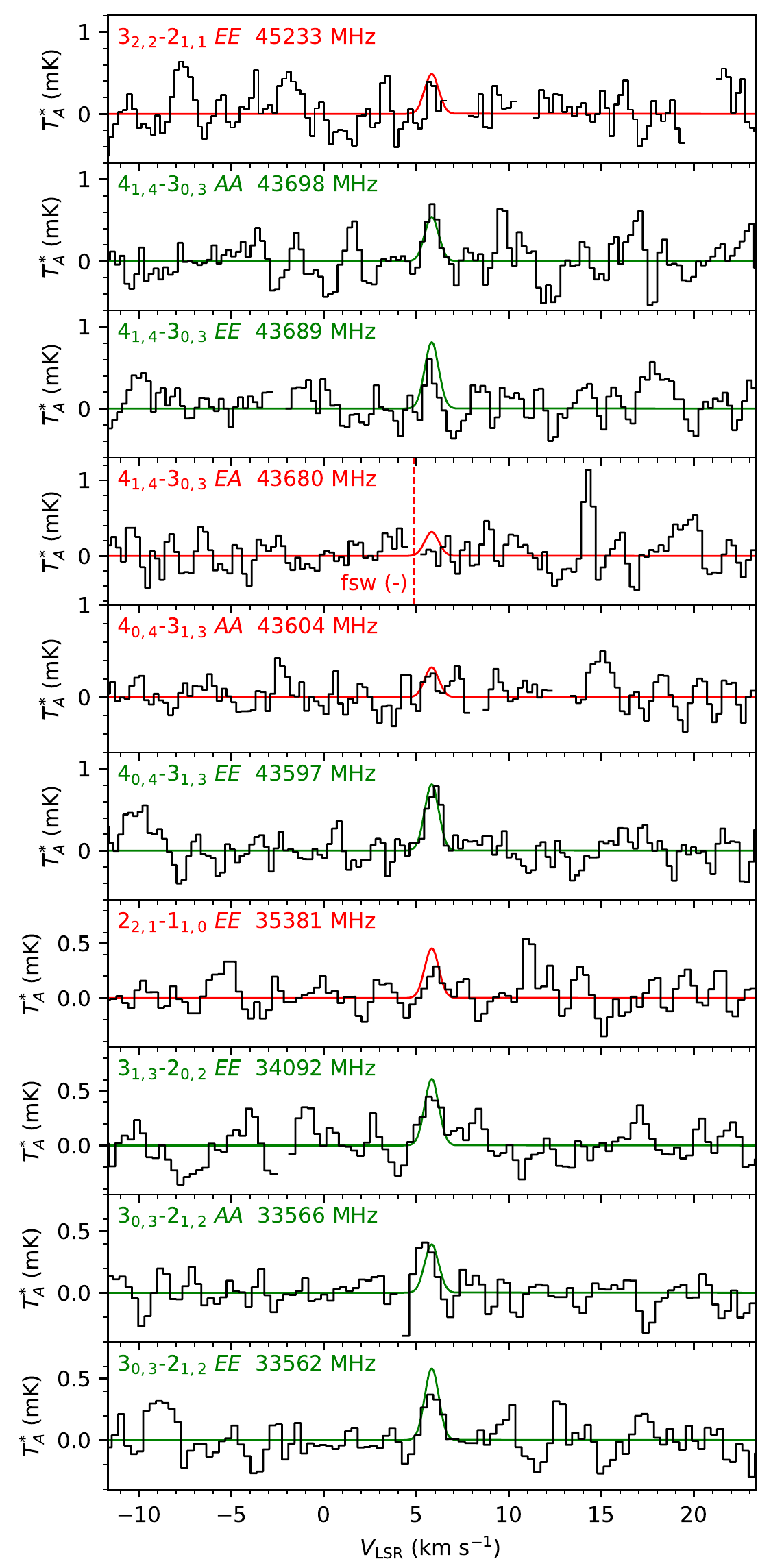}
\caption{Lines of CH$_3$COCH$_3$ observed in \mbox{TMC-1} are shown from bottom to top in order of increasing frequency (see Table~\ref{table:lines}). The panels in green show the lines that are well detected, while those in red show lines that are not clearly detected due to a variety of reasons (see text and caption in Table~\ref{table:lines}). The position of a negative frequency switching artifact is indicated by a dashed vertical line and labeled as fsw (-). The green/red solid lines correspond to the calculated spectra for $N$\,=\,1.4\,$\times$\,10$^{11}$\,cm$^{-2}$, $T_{\rm rot}$\,=\,6.0\,K, FWHM\,=\,0.90 km s$^{-1}$, and $\theta_s$\,=\,80\,$''$.} \label{fig:lines_ch3coch3}
\end{figure}

\subsection{Ethanol (C$_2$H$_5$OH)}

Ethanol has two conformers, $anti$ and $gauche$, depending on the orientation of the OH group. The most stable and the one reported here is the $anti$ form, the spectroscopy of which was taken from the Cologne Database for Molecular Spectroscopy \citep{Muller2005}\,\footnote{\texttt{https://cdms.astro.uni-koeln.de/}}, which in turn is based on \cite{Pearson2008} and \cite{Muller2016}. The dipole moment along the $b$ axis (all transitions observed here are $b$-type) is 1.438\,D \citep{Takano1968}.

We detected the four lines of $anti$ ethanol that are predicted to be the most intense in the Q band (see Table~\ref{table:lines} and Fig.~\ref{fig:lines_c2h5oh}). The three lines lying at 36417 MHz, 43026 MHz, and 46832 MHz are detected with very high signal-to-noise ratios (S/N), $\gtrsim$\,15, and are precisely centered around the systemic velocity of the source, $V_{\rm LSR}$\,=\,5.83 km s$^{-1}$ \citep{Cernicharo2020}. The line at 32742 MHz is also well detected although at a lower S/N and with an intensity lower than predicted by LTE. For a rotational temperature of 6.0 K, as derived below for C$_2$H$_5$OH, the predicted relative intensities for the four lines, in order of increasing frequency, are 0.80\,:\,0.50\,:\,1.00\,:\,0.68. From the observed velocity-integrated intensities (see Table~\ref{table:lines}), the resulting relative intensities are 0.33\,:\,0.55\,:\,1.00\,:\,1.03. The most puzzling point is that the first line should be at least twice more intense than it is observed. We do not have a fully satisfactory explanation for this point. It could happen that some weak line lying at $\pm$\,8 MHz and/or $\pm$\,10 MHz could produce a frequency switching negative artifact at the position of the 32742 MHz line, decreasing its intensity. Alternatively, non-LTE excitation effects may be playing a role. In spite of this puzzling issue on the relative intensity of the 32742 MHz line, we consider that the detection of ethanol in \mbox{TMC-1} should be secure because it would be very unlikely to have four unidentified lines that by coincidence lie at the precise frequencies of the four strongest transitions of C$_2$H$_5$OH. We note that the only previous detection of ethanol in a cold source, reported toward L483 by \cite{Agundez2019}, relied on one single line.

From a rotation diagram using the four lines of ethanol detected in \mbox{TMC-1} we derive a rotational temperature ($T_{\rm rot}$) of 6.0\,$\pm$\,0.8 K and a column density of (1.1\,$\pm$\,0.3)\,$\times$\,10$^{12}$ cm$^{-2}$, where we assumed a circular emission distribution with a diameter $\theta_s$\,=\,80$''$ \citep{Fosse2001}. The calculated line profiles are shown in Fig.~\ref{fig:lines_c2h5oh}, where we adopted as line width the arithmetic mean of the observed $\Delta v$ values.

\subsection{Acetone (CH$_3$COCH$_3$)}

Acetone is an asymmetric rotor in which the large amplitude internal motion of the two equivalent methyl groups leads to level splitting into $AA$, $EE$, $EA$, and $AE$ substates \citep{Groner2002}, as occurs for dimethyl ether. We adopted the spectroscopy from the Jet Propulsion Laboratory (JPL) catalogue \citep{Pickett1998}\,\footnote{\texttt{https://spec.jpl.nasa.gov/}}. The geometry of the molecule results in a nonzero dipole moment only along the $b$ axis, with a measured value of 2.93\,D \citep{Peter1965}.

We computed the line intensities of CH$_3$COCH$_3$ in LTE adopting a rotational temperature of 6.0 K, as derived for C$_2$H$_5$OH, and focused on the ten lines predicted to be the most intense in the Q band (see Table~\ref{table:lines} and Fig.~\ref{fig:lines_ch3coch3}). The five most intense predicted lines within this group (those with $T_{b, \rm \, calc}$\,$>$\,1.0 mK; see Table~\ref{table:lines}) are all detected with low to moderate S/N, in the range 5-9\,$\sigma$. Among the next five lines, predicted with $T_{b, \rm \, calc}$ in the range 0.6-1.0 mK (see Table~\ref{table:lines}), we only detected clearly one line, the one lying at 33566 MHz. There is another line at 35381 MHz predicted with a similar intensity which is detected only marginally (see Fig.~\ref{fig:lines_ch3coch3}). The three remaining lines lie above 40 GHz, where the sensitivity of the QUIJOTE data is worse than at lower frequencies. None of these three lines are clearly detected, although the observed spectra is consistent with the expected line intensities. For example, the lines at 43604 MHz and 45233 MHz are predicted with an intensity within the noise level of the data (see Fig.~\ref{fig:lines_ch3coch3}), while the line at 43680 MHz is not seen because there is a frequency-switching negative artifact which lies very close to the expected position of this line (see Fig.~\ref{fig:lines_ch3coch3}). In summary, among the set of ten lines predicted to be the most intense, we detected six lines with low to moderate S/N, while the four remaining lines are not detected because insufficient sensitivity or overlap with a negative frequency-switching artifact. From our experience in the detection of molecules through weak lines with the QUIJOTE data (e.g., \citealt{Agundez2021}), we consider that the detection of CH$_3$COCH$_3$ is secure because it would be very unlikely to have six lines of other species precisely centered at the frequencies of the strongest lines of acetone.

The six detected lines cover a low range of upper level energies and their velocity-integrated intensities have non negligible errors, which makes it difficult to precisely determine the rotational temperature of CH$_3$COCH$_3$. We therefore adopted a rotational temperature of 6.0 K, as determined for C$_2$H$_5$OH, and a source size with a diameter of 80$''$, as adopted for C$_2$H$_5$OH, and derived a column density of (1.4\,$\pm$\,0.6)\,$\times$\,10$^{11}$ cm$^{-2}$ for CH$_3$COCH$_3$. The calculated line profiles, adopting as line width the average of the $\Delta v$ values observed, are shown in Fig.~\ref{fig:lines_ch3coch3}.

\begin{figure}
\centering
\includegraphics[angle=0,width=\columnwidth]{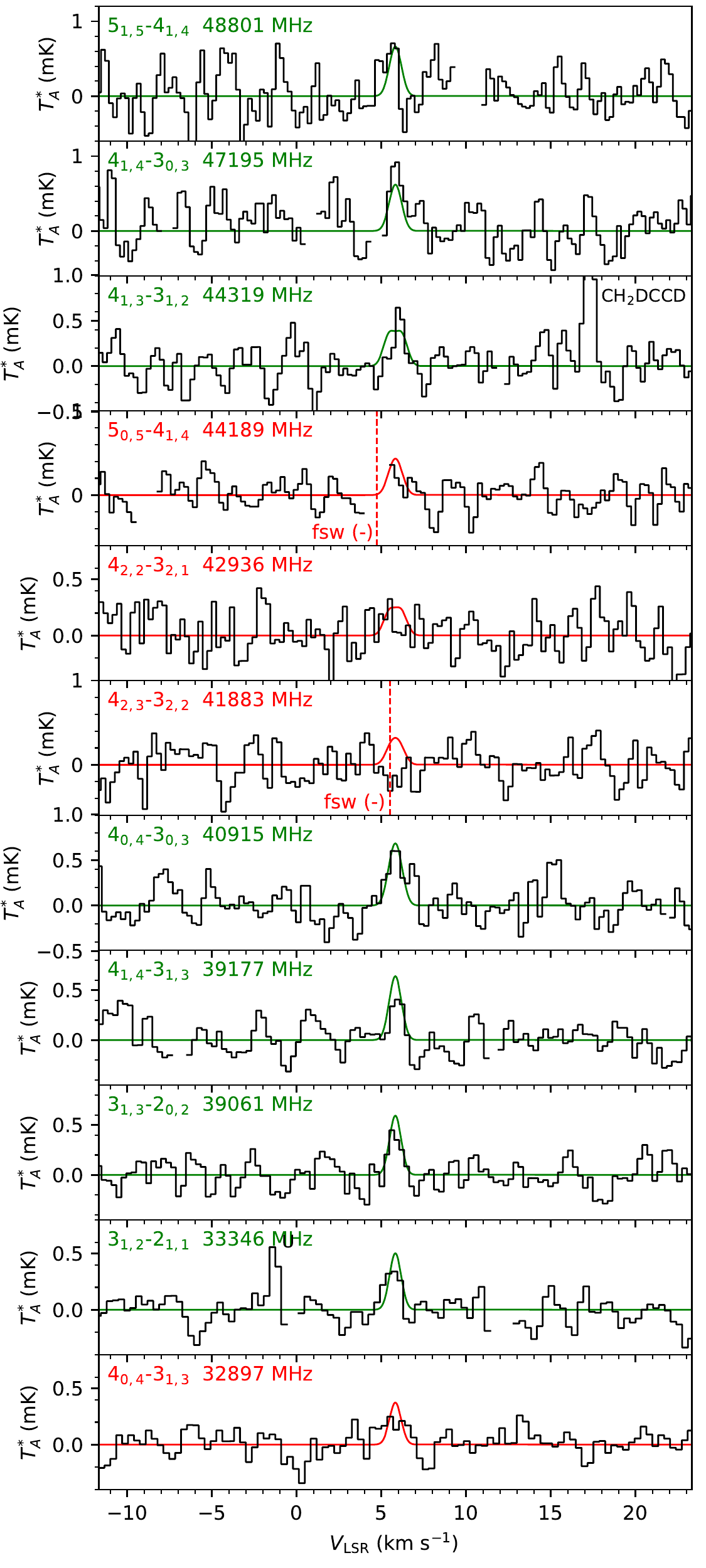}
\caption{Same as Fig.~\ref{fig:lines_ch3coch3} but for C$_2$H$_5$CHO, i.e. detected lines in green and non detected lines in red (see line parameters in Table~\ref{table:lines}). The green/red solid lines correspond to the calculated spectra for $N$\,=\,1.9\,$\times$\,10$^{11}$\,cm$^{-2}$, $T_{\rm rot}$\,=\,6.0\,K, FWHM\,=\,0.81 km s$^{-1}$, and $\theta_s$\,=\,80\,$''$.} \label{fig:lines_c2h5cho}
\end{figure}

\subsection{Propanal (C$_2$H$_5$CHO)}

Propanal has two conformers, $syn$ and $gauche$, depending on the orientation of the CHO group. The most stable one is the $syn$ form, which is the one detected in \mbox{TMC-1}. The rotational spectroscopy was taken from the CDMS \citep{Muller2005}, which is mostly based on \cite{Zingsheim2017}. The components of the dipole moment along the $a$ and $b$ axes are 1.71\,D and 1.85\,D, respectively \citep{Butcher1964}. Both $a$- and $b$-type transitions are observed here.

Similarly to the case of acetone, we computed the line intensities in LTE of $syn$ propanal assuming a rotational temperature of 6.0 K and focused on the lines expected with a brightness temperature $T_{b, \rm \, calc}$ above 0.7 mK. The eleven resulting lines are given in Table~\ref{table:lines} and shown in Fig.~\ref{fig:lines_c2h5cho}. There are eight lines within this selected group which are predicted with $T_{b, \rm \, calc}$\,$>$\,0.9 mK, from which we detected seven lines with low to moderate S/N in the range 5-8\,$\sigma$. The remaining line at 44189 MHz overlaps with a negative artifact resulting from the frequency-switching technique (see Fig.~\ref{fig:lines_c2h5cho}). The other four lines predicted to be less intense, with $T_{b, \rm \, calc}$ in the range 0.7-0.9 mK are not clearly detected for different reasons. The line at 32897 MHz is only marginally detected, while the one at 41883 MHz overlaps with a frequency-switching negative artifact, and the last one at 42936 MHz is predicted with an intensity within the noise level of the data. Even if the lines of propanal are detected with relatively low S/N, we are confident about the detection due to the high number of detected lines (seven) and because we can explain those lines which are not detected on the basis of insufficient sensitivity of overlap with negative artifacts from the frequency-switching observing mode. As in the case of acetone, it will be very unlikely to have seven lines from different species located at the precise frequencies of the most intense lines of $syn$ propanal.

As in the case of acetone, the sizable errors in the velocity-integrated intensities do not allow to constrain precisely the rotational temperature of propanal. We therefore adopted a rotational temperature of 6.0 K, as determined for C$_2$H$_5$OH. As with the previous molecules, we adopt an emission distribution with a diameter of 80$''$. The column density derived for C$_2$H$_5$CHO is (1.9\,$\pm$\,0.7)\,$\times$\,10$^{11}$ cm$^{-2}$. In Fig.~\ref{fig:lines_c2h5cho} we show the line profiles computed, where we adopted as FWHM the mean of the $\Delta v$ values observed.

\section{Chemical model}

\begin{figure*}
\centering
\includegraphics[angle=0,width=0.36\textwidth]{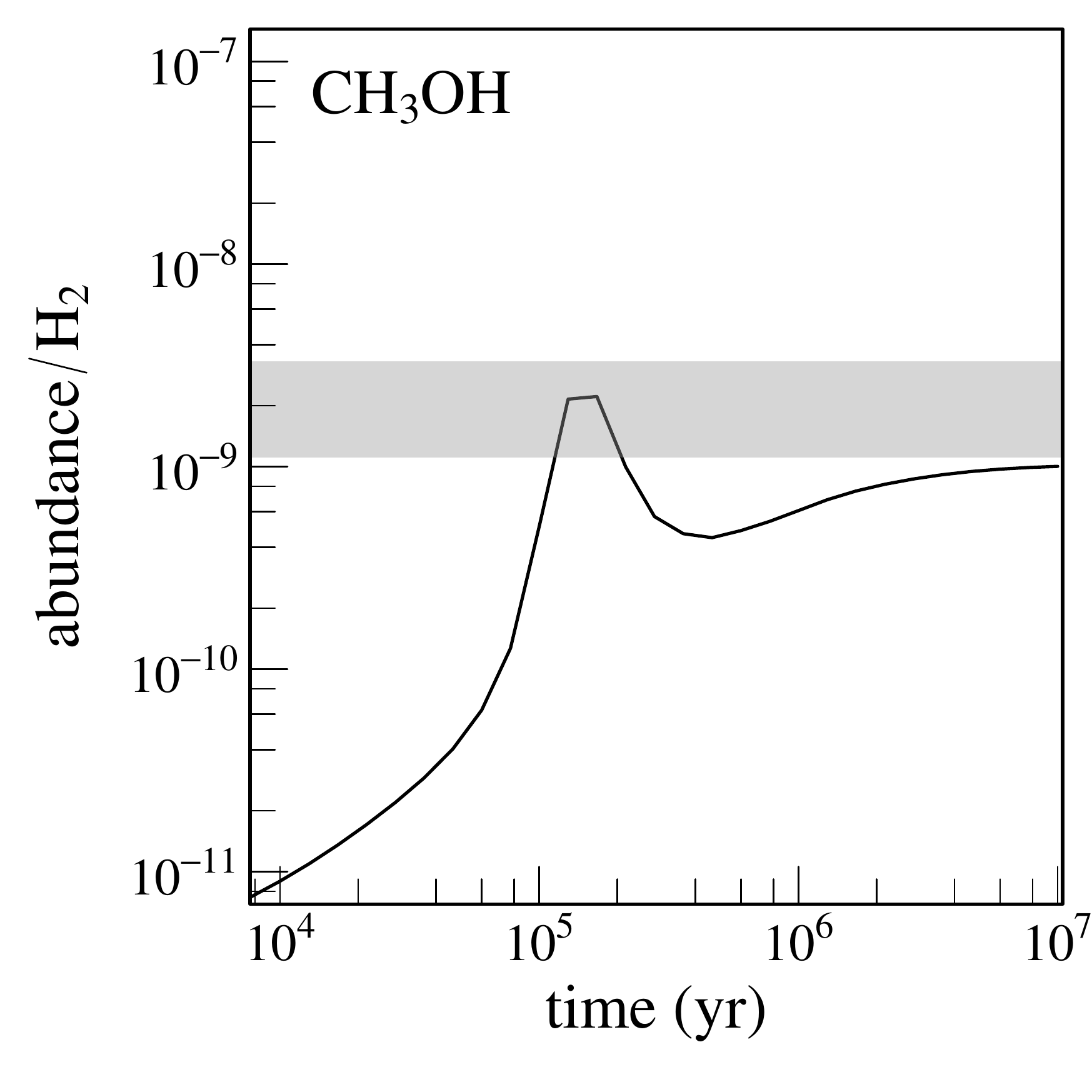} \hspace{1cm} \includegraphics[angle=0,width=0.36\textwidth]{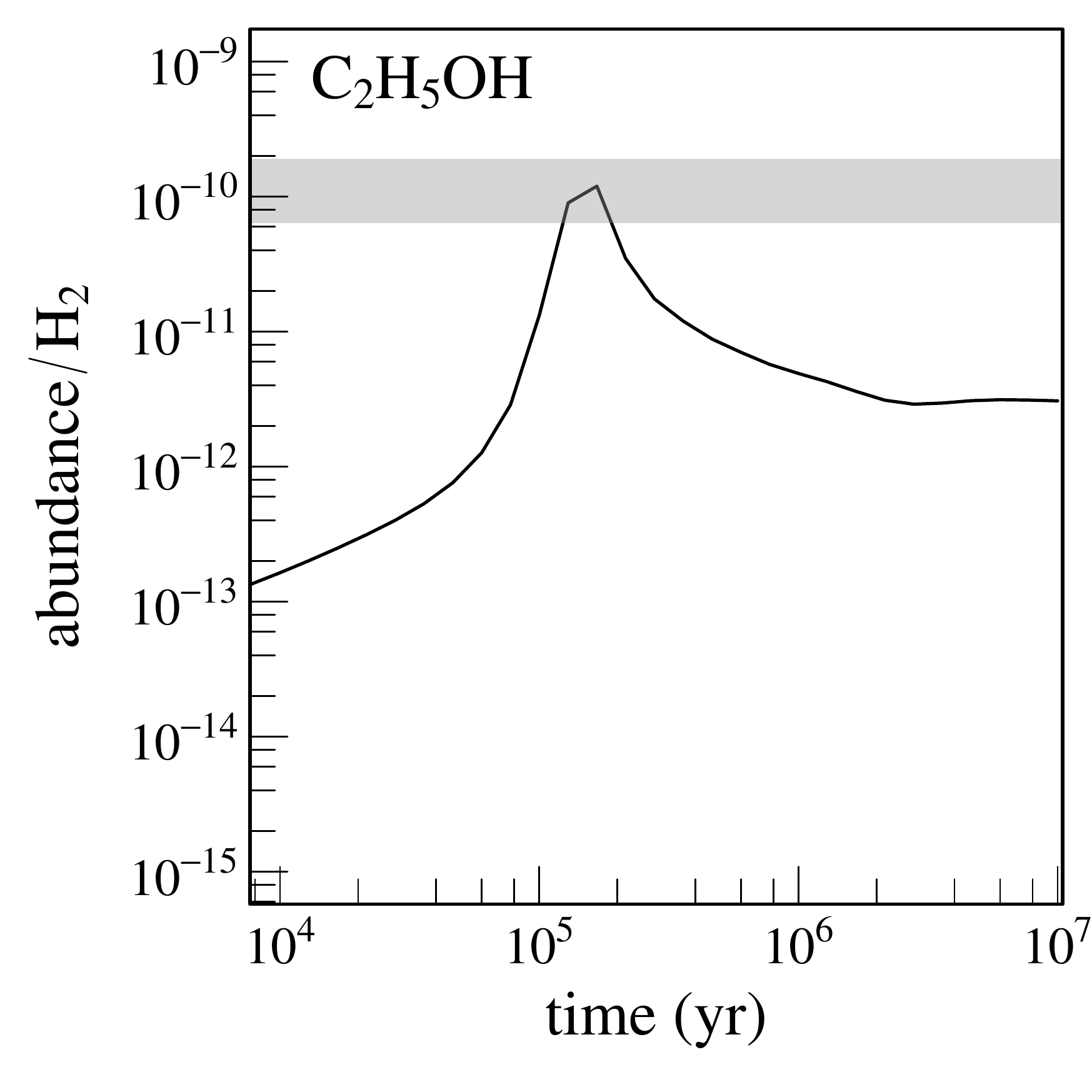} \includegraphics[angle=0,width=0.36\textwidth]{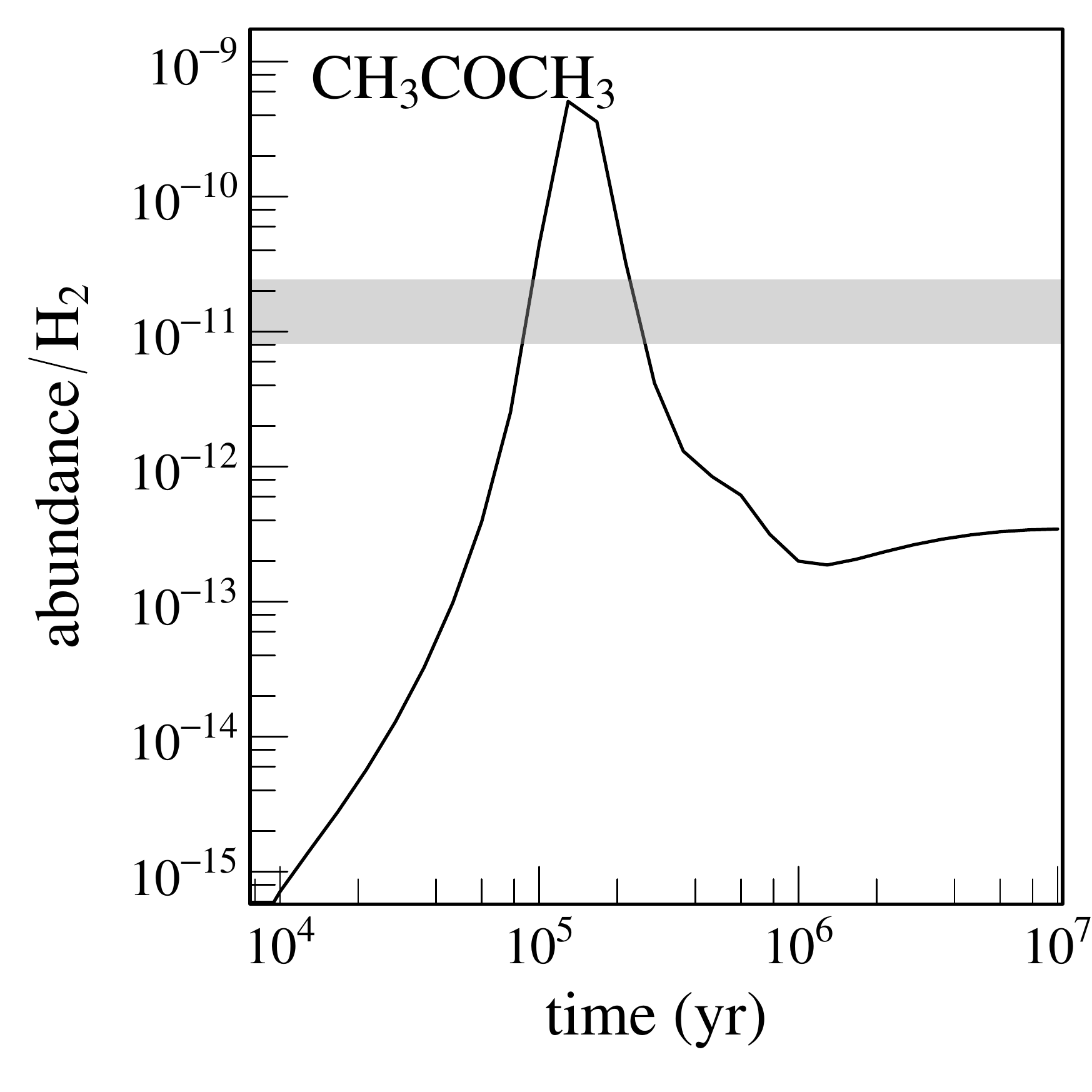} \hspace{1cm} \includegraphics[angle=0,width=0.36\textwidth]{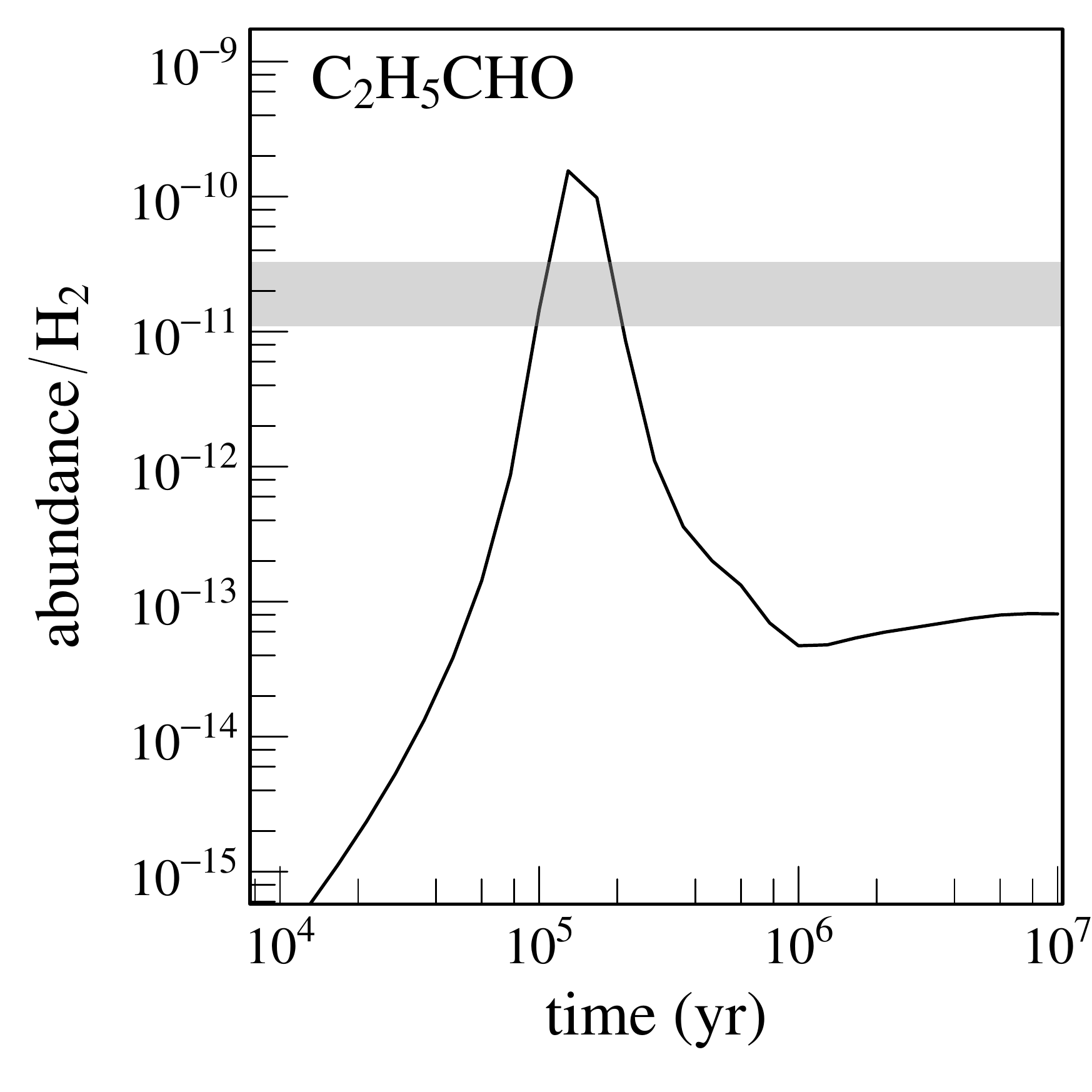}
\caption{Abundances of CH$_3$OH, C$_2$H$_5$OH, CH$_3$COCH$_3$, and C$_2$H$_5$CHO as a function of time calculated with our chemical model. The horizontal grey rectangles represent the abundances observed in \mbox{TMC-1} assuming an uncertainty of a factor of three.}\label{fig:abun}
\end{figure*}

To describe the chemistry of O-bearing COMs in \mbox{TMC-1} we used the Nautilus code, which is a three-phase (gas, dust grain ice surface, and dust grain ice mantle) time-dependent chemical model \citep{Ruaud2016}. The code has been recently updated with a better description of the chemistry of COMs in the gas phase and on grains \citep{Manigand2021,Coutens2022} and with the inclusion of sputtering of ices by cosmic-rays \citep{Wakelam2021}. To describe the physical conditions in \mbox{TMC-1} we use an homogeneous cloud with an H$_2$ volume density of 2.5\,$\times$\,10$^4$ cm$^{-3}$, a temperature of 10 K for both gas and dust, a visual extinction of 30 mag, and a cosmic-ray ionization rate of H$_2$ of 1.3\,$\times$\,10$^{-17}$ s$^{-1}$. All elements are assumed to be initially in atomic form, except for hydrogen, which is entirely molecular. The initial abundances are those of Table 1 of \cite{Hincelin2011}, the C/O elemental ratio being equal to 0.7 in that study.

The calculated abundances relative to H$_2$ for the O-bearing COMs detected in this study are shown in Fig.~\ref{fig:abun}. The observed abundances are relatively well reproduced by the model for a cloud age around 2\,$\times$\,10$^5$ yr. It should be noted that although the destruction pathways of COMs are fairly well constrained, this is not the case for the formation pathways. Destruction of COMs occurs mainly through reactions with H$_3^+$ and atomic carbon. The reactions with H$_3^+$ are an important destruction pathway for COMs because not only is the protonated form usually a minor product \citep{Lee1992}, but also the electronic dissociative recombination (DR) of the protonated form gives back very little of the original COM. For example, the DR of C$_2$H$_5$OH$_2^+$ produces less than 7\% of C$_2$H$_5$OH \citep{Hamberg2010}. The reactions with C are also an important destruction pathway for COMs because atomic carbon seems to react without a barrier with numerous COMs \citep{Husain1999,Shannon2014, Hickson2021}. The other important destruction pathways are the reactions with C$^+$, H$^+$, S$^+$, and He$^+$. With regard to the formation of O-bearing COMs in cold molecular clouds, there are several very different possible pathways.

First, bimolecular reactions between atomic oxygen and hydrocarbon radicals can be an important source of O-bearing COMs. Since atomic oxygen is very abundant, this mechanism can be very efficient when the hydrocarbon radical is present with a high enough abundance. Acetone is produced through the O + 2-C$_3$H$_7$ (CH$_3$CHCH$_3$) reaction and propanal is produced through the O + 1-C$_3$H$_7$ (CH$_2$CH$_2$CH$_3$) reaction \citep{Tsang1986,Hoyermann1979}. In our present version of the model we consider that C$_3$ does not react with oxygen atoms \citep{Woon1996}, which results in high abundances of C$_3$ derivatives in the ice \citep{Hickson2016} and a notable abundance for the C$_3$H$_7$ radical in the gas phase through chemical desorption \citep{Garrod2007,Minissale2016} and cosmic-ray sputtering \citep{Wakelam2021}. This is the main formation pathway for C$_2$H$_5$CHO and an important one for CH$_3$COCH$_3$. 

Second, radiative association between neutral radicals can directly produce O-bearing COMs. For example, the association of CH$_3$ and CH$_2$OH can yield C$_2$H$_5$OH, while CH$_3$ + CH$_3$CO can form CH$_3$COCH$_3$. These two reactions have not been studied to the best of our knowledge but are likely to be fast by comparison with CH$_3$ + CH$_3$O $\rightarrow$ CH$_3$OCH$_3$ + $h \nu$ \citep{Tennis2021}. However, these reactions are not very efficient in our \mbox{TMC-1} model because the calculated abundances of CH$_2$OH and CH$_3$CO are too small. It would be interesting to search for CH$_2$OH, a species whose microwave spectroscopy has been recently studied \citep{Bermudez2017, Chitarra2020,Coudert2022} to see if its abundance is significantly higher than calculated, which could increase the role of radiative associations between neutrals in the synthesis of COMs, as suggested by \cite{Balucani2015}.

The third way to produce O-bearing COMs is the DR of their protonated form. However, there are relatively few reactions producing the protonated form of the COMs, mostly ion-neutral radiative associations which compete with the faster proton transfer channel when this is exothermic. We used the reactions and rate coefficients from \cite{Herbst1987} and \cite{Herbst1990} to produce C$_2$H$_5$CHO and CH$_3$COCH$_3$, reviewing the different pathways to (CH$_3$)$_2$COH$^+$, in addition to the one postulated by \cite{Herbst1990}, CH$_3^+$ + CH$_3$CHO (see Appendix~\ref{appendix}). In any case, these routes play a secondary role in the formation of the O-bearing COMs observed in this study except for CH$_3$COCH$_3$, in which case the pathway initiated by the reaction OH + C$_3$H$_7^+$ is favored by the important production of C$_3$ and its derivatives.

The last possible way of production of COMs is through their synthesis on grains followed by desorption. As the majority of species are not mobile on grains at 10 K, except for atomic hydrogen and to a lesser extent atomic nitrogen, this requires an abundant precursor in the gas phase and an efficient low temperature desorption mechanism, such as chemical desorption \citep{Garrod2007,Minissale2016} or cosmic-ray sputtering \citep{Wakelam2021}. This is the case for methanol, formed from the surface hydrogenation of CO produced in the gas phase. This is also potentially the case for C$_3$ derivatives (methylacetylene, propene, propane) if C$_3$ does not react with atomic oxygen \citep{Woon1996,Hickson2016}, but not for COMs in general. However, since atomic carbon reacts without barrier with species on grains such as H$_2$CO \citep{Husain1999} and CH$_3$OH \citep{Shannon2014}, the reactive sticking of carbon atom to the grains induces the formation of COMs through an Eley-Rideal mechanism, as discussed by \cite{Ruaud2015}. In particular, in our model C$_2$H$_5$OH is mainly formed by the insertion of the C atom into the C$-$O bond of CH$_3$OH followed by hydrogenation.

In summary, our chemical model indicates that C$_2$H$_5$OH is mostly produced on grains, C$_2$H$_5$CHO is mainly formed by the reaction O + 1-C$_3$H$_7$, where 1-C$_3$H$_7$ itself is produced on grains, while CH$_3$COCH$_3$ is produced by the O + 2-C$_3$H$_7$ reaction (2-C$_3$H$_7$ itself is produced on grains) and the DR of (CH$_3$)$_2$COH$^+$, where (CH$_3$)$_2$COH$^+$ is produced by the reactions OH + C$_3$H$_7^+$, H$_2$O + C$_3$H$_5^+$, and C$_2$H$_4$ + H$_2$COH$^+$. The agreement between the observations and the simulations for CH$_3$COCH$_3$ and C$_2$H$_5$CHO is very dependent on the efficiency of the O + C$_3$ reaction and the efficiency of the desorption mechanisms. It is also indirectly dependent on the rate of the reactions of C$_3$H$_3^+$ and C$_3$H$_5^+$ with H$_2$, which are very slow at room temperature \citep{Lin2013} but could be open at low temperature due to tunneling. These hydrogenation reactions are essential to reproduce the observations of CH$_3$CCH \citep{Markwick2002} and C$_3$H$_6$ \citep{Marcelino2007}, which cannot be explained by the desorption of these species alone from the grains. It is important to emphasize the role of the O + C$_3$ reaction which, if slow, allows the observed O-bearing COM abundances to be reproduced but induces a very high abundance of C$_3$ in the gas phase which seems incompatible with the $^{13}$C fractionation of $c$-C$_3$H$_2$ \citep{Loison2020} and the abundances of CH$_3$CCH and C$_3$H$_6$ around protostars \citep{Manigand2021}.

An interesting test of the ability of our chemical model to explain the chemistry of O-bearing COMs in \mbox{TMC-1} is to verify if the model can account for the observed abundance of methanol. Even if the production mechanisms of methanol, ethanol, acetone, and propanal are different, the chemistry on grain surfaces is crucial for these species. It is essential for CH$_3$OH, C$_2$H$_5$OH, and C$_2$H$_5$CHO, and important for CH$_3$COCH$_3$. For the three latter molecules grain surface chemistry plays a role because C$_2$H$_5$OH is formed from the reaction of atomic carbon with adsorbed CH$_3$OH, while CH$_3$COCH$_3$ and C$_2$H$_5$CHO are formed from the desorption of C$_3$H$_x$ species followed by gas-phase chemistry. Similarly, the formation of methanol is essentially done on grains (more than 99\,\%) by the successive hydrogenation of CO, with part of this methanol being injected into the gas phase by two mechanisms: desorption induced by cosmic-rays \citep{Wakelam2021} and chemical desorption \citep{Minissale2016}. It is worth to note that the formation pathway to methanol in the gas phase from the dissociative recombination of protonated methanol is globally inefficient. On the one hand, the formation of CH$_3$OH$_2^+$ itself through either the radiative association between CH$_3^+$ and H$_2$O \citep{Gerlich2006} or the reaction between H$_3^+$ and HCOOCH$_3$ \citep{Lawson2012} is not very efficient. On the other hand, the production of methanol is a minor channel in the dissociative recombination of CH$_3$OH$_2^+$ with electrons \citep{Geppert2006}. The calculated abundance of CH$_3$OH (see Fig.~\ref{fig:abun}) shows a good agreement with the value derived from observations of \mbox{TMC-1} \citep{Pratap1997,Soma2015,Soma2018} for a cloud age around 2\,$\times$\,10$^5$ yr, as for the other O-bearing COMs detected in this study.

\section{Conclusions}

We detected, for the first time in a starless core (\mbox{TMC-1}), three new O-bearing COMs: ethanol, acetone, and propanal. These detections enlarge the inventory of O-bearing COMs known to be present in cold interstellar clouds, where previously other molecules of this type, such as methyl formate, dimethyl ether, propenal, and vinyl acohol have been found. These molecules are present at abundance levels of 10$^{-11}$-10$^{-10}$ with respect to H$_2$ in \mbox{TMC-1}. A chemical model including gas and dust chemistry is able to explain the formation of ethanol, acetone, and propanal in \mbox{TMC-1} based on a combination of grain chemistry, non-thermal desorption, and gas-phase chemical processes.

\begin{acknowledgements}

We acknowledge funding support from Spanish Ministerio de Ciencia e Innovaci\'on through grants PID2019-106110GB-I00, PID2019-107115GB-C21, and PID2019-106235GB-I00 and from the European Research Council (ERC Grant 610256: NANOCOSMOS). J.-C.\,L., K.\,H., and V.\,W. acknowledge the CNRS program ''Physique et Chimie du Milieu Interstellaire'' (PCMI) co-funded by the Centre National d\'Etudes Spatiales (CNES). We thank the referee for a critical reading of the article.

\end{acknowledgements}

\appendix

\section{Ionic formation pathways of CH$_3$COCH$_3$ and C$_2$H$_5$CHO} \label{appendix}

\cite{Herbst1990} proposed that acetone may be produced by the dissociative recombination of the ion (CH$_3$)$_2$COH$^+$,
\begin{equation}
\nonumber \rm (CH_3)_2COH^+ + e^- \rightarrow (CH_3)_2CO + H,
\end{equation}
where (CH$_3$)$_2$COH$^+$ would be formed in the reaction
\begin{subequations} \label{reac:ch3+_ch3cho}
\small
\begin{align}
\nonumber \rm CH_3^+ + CH_3CHO & \rightarrow \rm (CH_3)_2COH^+          & \rm \Delta H = -430\,kJ\,mol^{-1} & \rm ~~~~6\,\% \\
\nonumber                                          & \rightarrow \rm C_2H_5^+ + H_2CO    & \rm \Delta H = -132\,kJ\,mol^{-1} & \rm ~~~~55\,\% \\
\nonumber                                          & \rightarrow \rm H_2COH^+ + C_2H_4 & \rm \Delta H = -159\,kJ\,mol^{-1} & \rm ~~~~36\,\% \\
\nonumber                                          & \rightarrow \rm CH_3CO^+ + CH_4     & \rm \Delta H = -333\,kJ\,mol^{-1} & \rm ~~~~3\,\%,
\end{align}
\end{subequations}
for which the rate coefficient and branching ratios were measured at room temperature. The measured rate coefficient for the three-body association was used to calculate the rate coefficient of the radiative association at low temperature using phase space theory and assuming an exit barrier to the production of bimolecular products (C$_2$H$_5^+$ + H$_2$CO, H$_2$COH$^+$ + C$_2$H$_4$, and CH$_3$CO$^+$ + CH$_4$).

\begin{figure*}[hb!]
\centering
\includegraphics[angle=0,width=0.74\textwidth]{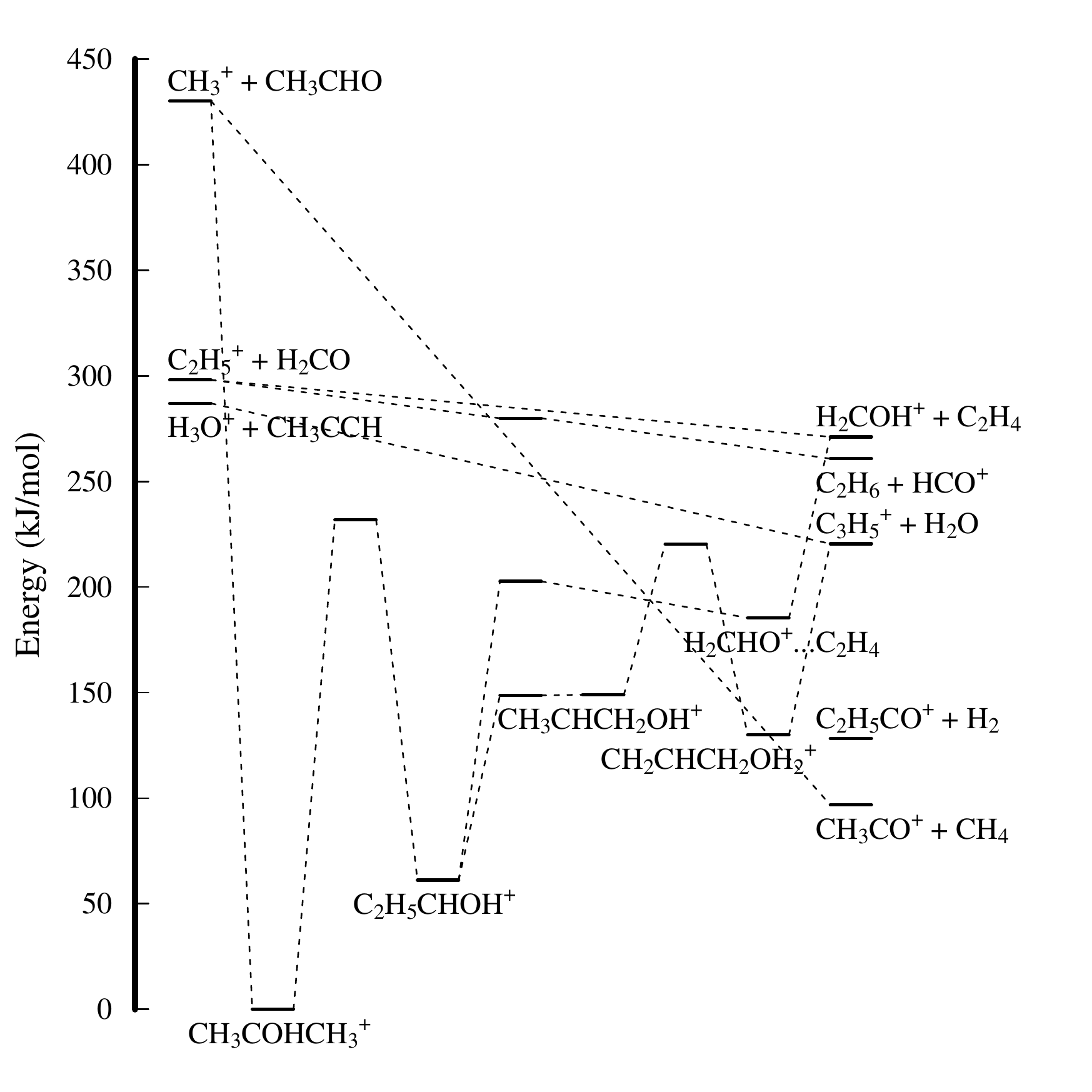}
\caption{Schematic representation of the PES of (CH$_3$)$_2$COH$^+$.} \label{fig:pes}
\end{figure*}

To explore other possible (CH$_3$)$_2$COH$^+$ formation pathways we performed various theoretical calculations on the global (CH$_3$)$_2$COH$^+$ potential energy surface (PES) using density functional theory (DFT) employing the M06-2X functional \citep{Zhao2008a} coupled with the aug-cc-pVTZ (AVTZ) basis set. This functional has been shown to have good accuracy for the prediction of main group thermochemistry and barrier heights \citep{Zhao2008b}. The results of our calculations are shown in Fig.~\ref{fig:pes}.

It appears that the product channels C$_2$H$_5^+$ + H$_2$CO, H$_2$COH$^+$ + C$_2$H$_4$, C$_3$H$_5^+$ + H$_2$O and H$_3$O$^+$ + CH$_3$CCH are all barrierless, which implies a decrease of the theoretical rate coefficient of the radiative association channel with respect to the value of \cite{Herbst1990}, although it opens other possible pathways for the production of (CH$_3$)$_2$COH$^+$. For the C$_2$H$_5^+$ + H$_2$CO and H$_3$O$^+$ + CH$_3$CCH reactions, protonation was observed to proceed in the absence of competing channels \citep{Tanner1979,Milligan2002}. The CH$_3$CO$^+$ + CH$_4$ reaction involves large barriers and does not play a role in the formation of CH$_3$CO or C$_2$H$_5$CHO. The two reactions, in addition to CH$_3^+$ + CH$_3$CHO, that can produce the protonated forms of acetone and propanal through radiative association are H$_2$COH$^+$ + C$_2$H$_4$ and C$_3$H$_5^+$ + H$_2$O. The calculation of these rate coefficients is beyond the scope of this article but should be done in the future. Here we used values that seem realistic (a few 10$^{-10}$ cm$^3$ s$^{-1}$ at 10 K) and we also used the maximum rate coefficient calculated by \cite{Herbst1990}, 2\,$\times$\,10$^{-10}$ cm$^3$ s$^{-1}$ at 10 K, for the reaction 
\begin{equation}
\nonumber \rm CH_3^+ + CH_3CHO \rightarrow (CH_3)_2COH^+ + h\nu.
\end{equation}
However, even using these rate coefficients and considering large branching ratios (30\,\%) for the production of CH$_3$COCH$_3$ and C$_2$H$_5$CHO by dissociative recombination of their protonated forms with electrons, the production of CH$_3$COCH$_3$ and C$_2$H$_5$CHO is largely insufficient to reproduce the observations if their protonated forms are formed only by the radiative association reactions CH$_3^+$ + CH$_3$CHO, H$_2$COH$^+$ + C$_2$H$_4$, and C$_3$H$_5^+$ + H$_2$O.

An alternative route for the production of (CH$_3$)$_2$COH$^+$ is the OH + $i$-C$_3$H$_7^+$ reaction:
\begin{subequations} \label{reac:ch3+_ch3cho}
\small
\begin{align} 
\nonumber \rm OH + {\textit i}-C_3H_7^+ & \rightarrow \rm (CH_3)_2COH^+ + H    & \rm \Delta H = -125\,kJ\,mol^{-1} \\
\nonumber                                                & \rightarrow \rm CH_3CHOH^+ + CH_3 & \rm \Delta H = -99\,kJ\,mol^{-1}~~ \\
\nonumber                                                & \rightarrow \rm C_3H_6^+ + H_2O       & \rm \Delta H = -131\,kJ\,mol^{-1} \\
\nonumber                                                & \rightarrow \rm C_3H_5 + H_3O^+       & \rm \Delta H = -62\,kJ\,mol^{-1}~~
\end{align}
\end{subequations}
We studied this reaction at the M06-2X/AVTZ level and found by relaxing the geometry at each distance that the reaction producing the CH$_3$CHOHCH$_3^+$ adduct is barrierless. The evolution of this adduct can lead to different products including (CH$_3$)$_2$COH$^+$ + H, which is the most exothermic path with a very small barrier (or none at all) in the exit channel and therefore most likely an important if not the main product channel.

\end{document}